\documentclass{ws-procs9x6}

\newcommand{\fth}{\frac{3}{2}}
\newcommand{\ffh}{\frac{5}{2}}

\begin{document}

\title{Meson Production on the Nucleon in the Giessen K-Matrix Approach\footnote{\uppercase{W}ork partially
supported by \uppercase{FZ J}\"ulich}}

\author{H. Lenske, V. Shklyar, U. Mosel}

\address{Institut f\"ur Theoretische Physik, Universit\"at Giessen\\
Heinrich-Buff-Ring 16\\
D-35392 Giessen, Germany\\
E-mail: horst.lenske@physik.uni-giessen.de}

\maketitle

\abstracts{Meson production on the proton is described in a coupled
channels K-matrix approach. Both hadronic and photonic scenarios are
taken into account on the same theoretical footing. At tree-level
the Born amplitudes are obtained from an effective Lagrangian with
phenomenologically adjustable coupling constants. The spectral
distributions, resonances and their width and the background
contributions are obtained within the same approach, thus accounting
properly for interference effects. Applications to omega- and
associated strangeness production on the proton are discussed.}

\section{Introduction}

The nucleon as an entity of strongly interacting constituents is
playing the gateway to low-energy QCD as realized in hadrons.
Investigating its structure and dynamics by various probes and
measuring a variety of observables will complete the much wanted
data base on spectral properties of resonances, their excitations
and decays, including information about branching ratios into the
various meson-nucleon channels. An equally important motivation is
the search for {\em missing resonances} trying to bridge the gap
between the number of excited states of the nucleon predicted by
quark models and the -- at least until now -- much fewer resonances
seen in pion- or photon-induced reactions. In any respect, research
on the nucleon structure will necessarily include the need for a
good understanding of reaction dynamics. For that aim a realistic
description is necessary, accounting properly for the interplay of
various production channels, the interference among resonant and
non-resonant parts of the scattering amplitudes. Accepting this as a
guide line, it is obvious that single channel descriptions in
general will fail.

Coupled channel approaches are meeting these goals. They are able to
account for the various aspects of meson-nucleon and photon-nucleon
interactions and the {\em cross talk} between the various dynamical
sectors. Only with such a more involve description we will be able
to distinguish between the dynamical and the intrinsic, QCD related
properties of spectral structures observed in cross sections. The
Giessen model formulated a few years ago in
\cite{Feuster:98a,Feuster:98b} and extended subsequently in
\cite{Penner:02a,Penner:02b,Penner:02c} is in line with these
requirements. The model uses a unitary coupled-channel effective
Lagrangian approach. It has been successfully applied in the
analysis of pion- and photon-induced reactions in the energy region
up to 2 GeV. The resonance couplings are simultaneously constrained
by available experimental data from all open channels. While our
previous analyses \cite{Penner:02a,Penner:02b} have been restricted
to resonances with spin  $J\leq \fth$ we have extended the
description quite recently to higher spin states \cite{Shklyar:04},
thus enlarging the model space and increasing the predictive power
of the calculations. We now include essentially all channels
contributing significantly to the cross sections in the energy
region up to 2~GeV. The Giessen model is briefly summmarized in
sect. \ref{sec:theory}. We then discuss recent applications to
$\omega$-meson production in sect. \ref{sec:omega} and to the
associated strangeness production in sect. \ref{sec:KL}. The report
closes with a short summary and outlook in sect. \ref{sec:sumry}.

\section{The Giessen Model: Coupled Channels K-Matrix Description of Meson
Production}\label{sec:theory}

\begin{figure}[ht]
%\epsfxsize=10cm   %width of figure - will enlarge/reduce the figures
%\epsfbox{fig3.eps}
%\figurebox{2cm}{3cm}{} %to have a box alone
\centerline{\epsfxsize=4.1in\epsfbox{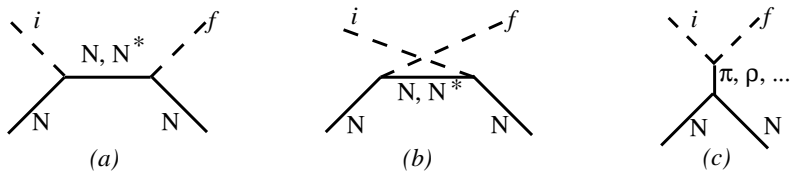}}
\caption{Born-diagrams in the s,u, and t channel contributing to the
Bethe-Salpeter equation. \label{fig1}}
\end{figure}

The Giessen model describes meson production on the nucleon on the
basis of an effective Lagrangian. At the energy scales considered
here the appropriate degrees of freedom are the nucleons and
hyperons from the basic SU(3) flavor octet and their excited states
and, on the meson side, the pseudoscalar and vector meson octet
states, supplemented by the photon and the electromagnetic coupling
of the hadrons \cite{Penner:02a}. The principal structure of the
model at tree-level is depicted in Fig.\ref{fig1}. Here, we only
briefly discuss the K-matrix part of the approach. The starting
point is the decomposition of a Green function into a principal
value and pole part given by a Dirac delta-function:
\begin{equation}\label{eq:Gmat}
G_{bs}\equiv \frac{P}{H-\omega}+i\pi\delta(H-\omega) \quad .
\end{equation}
With the propagators cast into this form the Bethe-Salpeter equation
can be represented by a set of two coupled equations
\begin{eqnarray}\label{eq:BSEQ}
\mathcal{K}=\mathcal{V}+\mathcal{V}\frac{P}{H-\omega}\mathcal{K}
\sim \mathcal{V} \\
\mathcal{M}=\mathcal{K}+i\mathcal{K}\delta(H-\omega)\mathcal{M}
\end{eqnarray}
where the first equation defines the K-matrix, corresponding for a
Hermitian Born amplitude $\mathcal{V}=\mathcal{V^\dag}$ to the real
part of the scattering amplitude. The solution of the above equation
for a multi-channel problem is a matrix containing the scattering
amplitudes from channels $\alpha$ to channels $\beta$
\begin{equation}\label{eq:Mab}
\mathcal{M}_{\alpha\beta}\equiv\left[\frac{\mathcal{K}}{1-i\mathcal{K}}
\right]_{\alpha\beta}\sim\left[\frac{\mathcal{V}}{1-i\mathcal{V}}
\right]_{\alpha\beta}
\end{equation}
where $\alpha$ and $\beta$ denote any of the photoproduction or
hadronic production channels. The validity and quality of the
K-matrix approach has been tested positively by various groups, e.g.
\cite{Sato:1996}.

\section{$\omega$ Meson Production off the Nucleon}\label{sec:omega}

In this section our primary interest is the $\omega$ meson
production in $\pi p$ and $\gamma p$ reactions, as discussed in
detail in \cite{Shklyar:05a}. Most of the theoretical studies of
this reaction are based on a rather simplified single channel
effective Lagrangian approach, e.g. \cite{Zhao:98,Titov:02}. But
there is agreement on the importance of the t-channel
$\pi_0$-exchange contributions, which were studied by Friman and
Soyeur \cite{Friman:96}. However, the claims by the various models
on the contributions of different resonances to the $\omega N$ final
state are controversial \cite{Shklyar:05a}. Compared to our previous
findings \cite{Penner:02a,Penner:02b} we observe significant changes
by inclusion of spin-$\ffh$ resonance contributions. To provide an
additional constraint on the resonance couplings to $\omega N$ we
also included the recent data on the spin density matrix obtained by
the SAPHIR group \cite{SAPHIR:03}.

\subsection{Hadronic Production: $\pi N \to \omega
N$}\label{ssec:hadronicomega}

All experimental data on  the $\omega$-meson production in the $\pi
N$ scattering have been measured before 1980 and therefore have
rather poor statistics. In total, there  are 115 data points which
includes differential and total cross sections data. The inclusion
of spin-$\ffh$ resonance contributions affects the $\omega N$ final
state considerably. The main contributions close to the threshold
come from the $P_{13}$ and $D_{15}$ partial waves. The resonance
part of the production amplitude is dominated by the $D_{15}(1675)$
state \cite{Shklyar:05a}. Overall, the inclusion of spin-$\ffh$
resonances shifts strength to the $P_{13}$ and $D_{15}$ partial
waves.

We also find strong contributions from the $P_{13}$ partial wave to
the $\pi N \to \omega N$ reaction what has been already reported in
\cite{Penner:02a}. The strength in this partial wave is shifted to
the lower energies and becomes more pronounced at the reaction
threshold. A peaking behavior seen in the $P_{13}$ partial cross
section is due to the interference pattern between $P_{13}$
resonances and background contributions to the $\omega N$ channel.
Hence, this is a representative example that the collaboration of
resonance and background features can produce structures in cross
sections which are easily misinterpreted as a resonance. Since the
major contributions to the $\pi N\to \omega N$ reaction come from
the $P_{13}$ and $D_{15}$ waves, it is interesting to look at the
$\pi N$ inelasticity for these partial waves which are found in
\cite{Shklyar:05a}. They lead to the conclusion that probably
inelasticities from other channels, e.g. ($3\pi N$), should also be
included.

\begin{figure}[ht]
  \begin{center}
{\includegraphics*[width=4.1in]{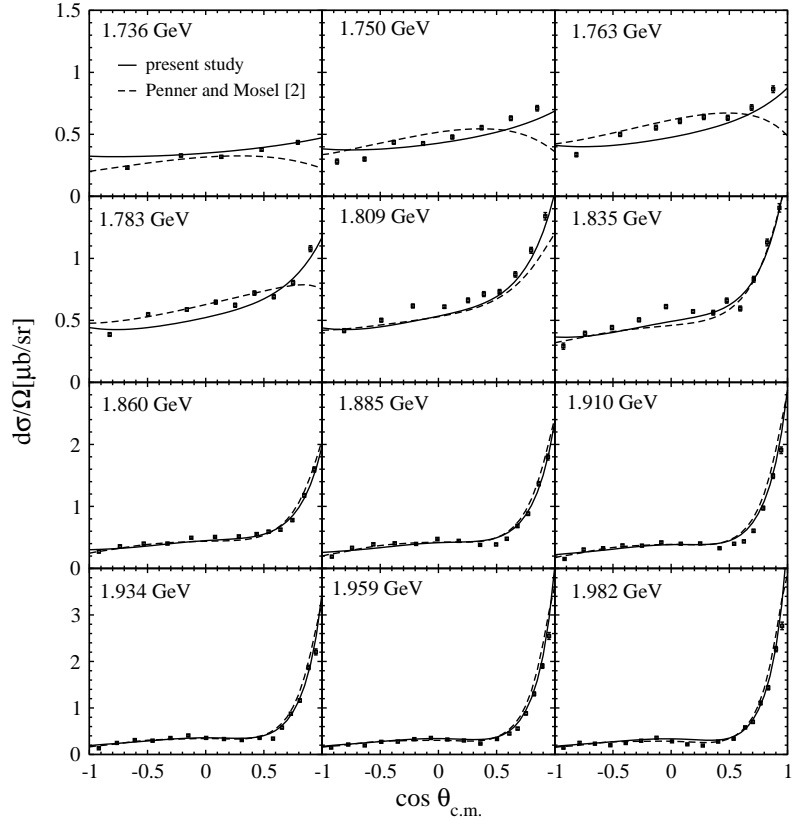}}
       \caption{
$\gamma N \to \omega N$ differential cross sections  in comparison
with the SAPHIR data \protect\cite{SAPHIR:03} and our previous
results from \protect\cite{Penner:02b}.
      \label{omgN_photo_dif}}
\end{center}
\end{figure}

\subsection{Photoproduction: $\gamma N \to \omega N$}\label{ssec:photo-omega}
The differential $\omega$ meson photoproduction  cross sections are
presented in Fig.\ref{omgN_photo_dif}. With the $\ffh$ components
included we obtain $\chi_{\gamma\omega}^2$=4.5 which significantly
improves our previous result ($\chi_{\gamma\omega}^2$=6.25)
\cite{Penner:02a,Penner:02b}. The strong $\pi^0$ exchange lead to a
peaking behavior of the calculated differential cross sections at
forward angles which are clearly visible in the SAPHIR measurements
\cite{SAPHIR:03} above 1.783 GeV and the theoretical cross sections,
both displayed in Fig. \ref{omgN_photo_dif}. More detailed
information of the production mechanism is obtained from observables
measuring the spin degree of freedom of the $\omega$ meson. In the
Gottfried-Jackson frame, where the initial photon and exchange
particle are in their rest frame, and $z$-axis is in the direction
of the incoming photon momentum, the calculation gives
$\rho_{00}^{GJ}=0$. The experimental value of $\rho_{00}^{GJ}$ for
forward directions, where the $\pi^0$ exchange dominates, was
measured by  SAPHIR and  found to be in the range of
$\rho_{00}^{GJ}=0.2\cdots 0.3$. Thus, the nonzero matrix element
testifies that even in this kinematical region other mechanisms
(rescattering effects, interference with resonances) must be
important.

\begin{figure}
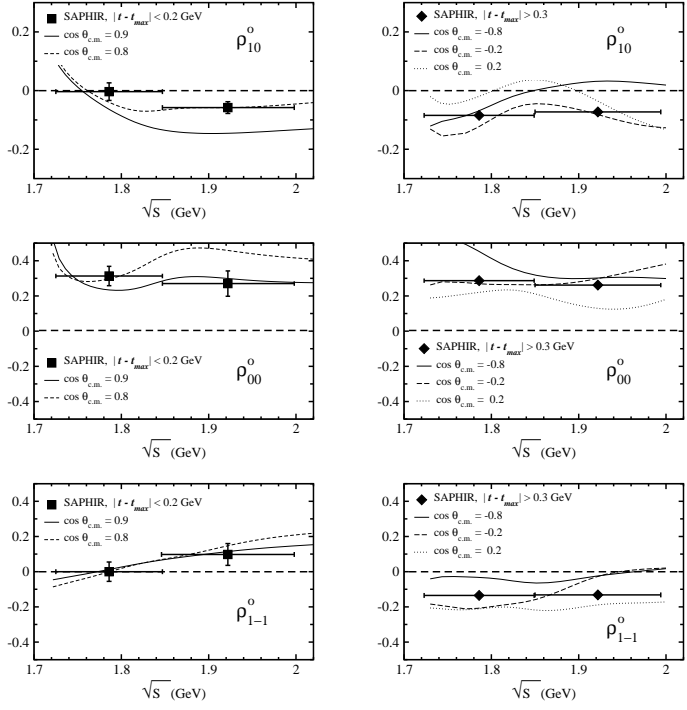

  \begin{center}
%{\includegraphics*[width=4.1in]{fig3.eps}}
    \parbox{10.25cm}{
      \parbox{4.8cm}{\includegraphics*[width=4.1cm]{fig3a.eps}}
      \parbox{4.6cm}{\includegraphics*[width=4.1cm]{fig3b.eps}}\\
      \parbox{4.8cm}{\includegraphics*[width=4.1cm]{fig3c.eps}}
      \parbox{4.6cm}{\includegraphics*[width=4.1cm]{fig3d.eps}}\\
      \parbox{4.8cm}{\includegraphics*[width=4.1cm]{fig3e.eps}}
      \parbox{4.6cm}{\includegraphics*[width=4.1cm]{fig3f.eps}}\\
       }
       \caption{Spin density matrix elements in the helicity frame compared to the
SAPHIR measurements \protect\cite{SAPHIR:03}. \label{omgN_spin}}
\end{center}
\end{figure}

Beside the $\pi^0$ exchange the largest contributions to $\omega$
meson photoproduction comes from the subthreshold spin-{$\ffh$}
resonances: $D_{15}(1675)$ and $F_{15}(1680)$. Since the $\pi^0$
exchange  above 1.8 GeV strongly influences the $\gamma N \to \omega
N$ reaction a consistent identification of individual resonance
contributions from only the partial wave decomposition is difficult.
The  $P_{13}(1900)$, and $F_{15}(2000)$, and $D_{13}(1950)$ states
which lie above the reaction threshold hardly influence the reaction
due to their small couplings to $\omega N$. Despite of the small
relative contribution from the $D_{15}$ and $F_{15}$ waves to the
$\omega$ photoproduction the cross sections are strongly affected by
spin-$\ffh$ states because of the destructive interference pattern
between the $\pi^0$ exchange and these resonance contributions.

While  $F_{15}(1680)$  plays only a minor role in the $\pi N \to
\omega N$ reaction the contribution from this state becomes more
pronounced in the $\omega$ meson photoproduction because of its
large $A^p_{\fth}$ helicity amplitude. The importance of the
$F_{15}(1680)$ resonance to the $\omega$ meson photoproduction was
also  found by Titov and Lee \cite{Titov:02} and by Zhao
\cite{Zhao:2000}. However, in contrast to \cite{Titov:02} where also
a large effect from $D_{13}(1520)$ was observed we do not find any
visible contribution from this state. In fact, as discussed in
\cite{Shklyar:05a} a strong contribution found in the $D_{13}$
partial wave, resembling a resonance structure, comes in fact from
non-resonant $\pi^0$ exchange.

The spin density matrix elements $\rho_{rr'}$ extracted from the
SAPHIR data \cite{SAPHIR:03} are an outcome of the averages over
rather wide energy and angle regions, see Fig. \ref{omgN_spin}. The
inclusion of measured $\rho_{rr'}$ into the calculations provides a
strong additional constraint on the relative partial wave
contributions and finally on the resonance couplings. A satisfactory
description of the spin density matrix is obtained in a wide energy
region. Since the $\rho_{rr'}$ data put strong constraints on the
$\gamma p \to \omega p$ reaction mechanism there is an urgent need
for  precise measurements of the spin density matrix in more narrow
energy bins to determine the reaction picture. Further details on
beam asymmetries are found in \cite{Shklyar:05a}.

\section{\label{sec:KL}Associated Strangeness Production on the Nucleon}
Since the recent  $K\Lambda$ photoproduction data
\cite{Glander:2003,McNabb:2003} give an indication for 'missing'
resonance  contributions, a combined analysis of the $(\pi,\gamma) N
\to K\Lambda$ reactions becomes inevitable to pin down these states.
Assuming small couplings to $\pi N$, these 'hidden' states should
not exhibit themselves in the pion-induced reactions and,
consequently, in the $\pi N \to K\Lambda $ reaction. The decay
ratios to the non-strange final states  and the electromagnetic
properties can be found in \cite{Shklyar:05a}. Our most recent
results in the extended approach are given in \cite{Shklyar:05b}. In
the that work, we have considered the partially contradicting CLAS
and SAPHIR data separately by performing independent fits to either
of the two data sets. In the following, the corresponding results
are denoted by the indices $C$ and $S$, respectively.

\subsection{\label{ssec:KL_hadr}Hadronic Strangeness Production: $\pi N \to K\Lambda$}

The $S$- and $C$- calculations differ in their description of the
non-resonance couplings to $K\Lambda$. As a consequence, different
background strengths are obtained for the $S_{11}$, $P_{11}$, and
$P_{13}$ partial waves while leaving the $P_{13}$(1720) and
$P_{13}(1900)$ resonance couplings almost unchanged
\cite{Shklyar:05b}. Comparing  the $S$- and $C$-parameter sets, the
largest difference in the resonance parameters  is observed for the
$P_{11}(1710)$ state. This resonance is found to be almost
completely of inelastic origin with a small branching ratio to ${\pi
N}$ \cite{Shklyar:05b}. This state gives only a minor contribution
to the reaction  and the observed  difference in the $P_{11}$
partial wave between $S$- and $C$-results is due to the Born term
and the $t$-channel exchange contributions.

The calculated differential cross sections corresponding to the $S$-
and $C$-coupling sets are found in \cite{Shklyar:05b}. Both results
show a good agreement with the experimental data in the whole energy
region. A difference between the two solutions is only found at
forward and backward scattering angles. This is due to the fact that
the CLAS photoproduction cross sections rise at backward angles
which is not observed by the SAPHIR group (see discussion below). At
other scattering angles the $S$ and $C$ results are very similar.
The differences between $S$- and $C$-calculations are more
pronounced for the $\Lambda$-polarization. Again, the main effect is
seen at the backward angles where the polarization changes its sign
in the $C$-calculations. Unfortunately, the quality of the data does
not allow to determine the reaction mechanism further.

\subsection{\label{ssec:KL_photo}Photoproduction of Strangeness: $\gamma N \to K\Lambda$}

The older SAPHIR measurements \cite{Tran:1998} show a resonance-like
peak in the total photoproduction cross section around 1.9 GeV. The
more recent SAPHIR \cite{Glander:2003} and  CLAS \cite{McNabb:2003}
data confirm the previous findings. However, the interpretation of
these data is controversial leaving open questions whether in these
measurements contributions from presently unknown resonances are
observed or if they can be explained by already established reaction
mechanisms.

\begin{figure}
  \begin{center}
{\includegraphics*[width=4.1in]{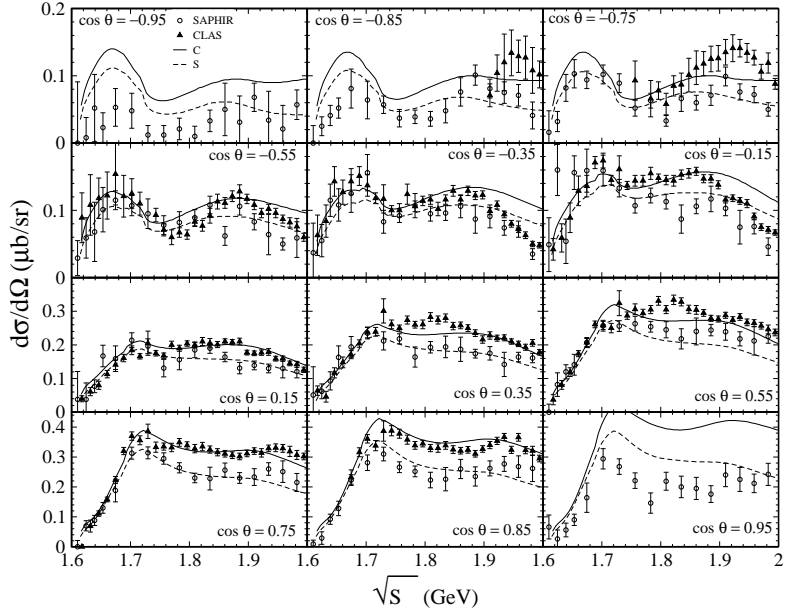}}
       \caption{
Comparison of the differential cross sections for the reaction
$\gamma p \to K^+\Lambda$ calculated using $C$ and $S$ parameter
sets. Experimental data are taken from
\protect\cite{McNabb:2003}(CLAS) and
\protect\cite{Glander:2003}(SAPHIR).
      \label{fig5:KL}}
  \end{center}
\end{figure}

Guided by the results of \cite{Penner:02b} we have performed a new
coupled-channel study of this reaction using separately the CLAS and
SAPHIR measurements as two independent input sets. The main
difference between the CLAS and SAPHIR data is seen at backward and
forward directions, Fig.~\ref{fig5:KL}. Both measurements show two
peaks but disagree in the absolute values of the corresponding
differential cross sections. Also, the second bump in the CLAS data
is shifted to the lower energy 1.8 GeV for the scattering angles
corresponding to $\cos\theta$=0.35 and $\cos\theta$=0.55.

\begin{figure}
  \begin{center}
{\includegraphics*[width=4.1in]{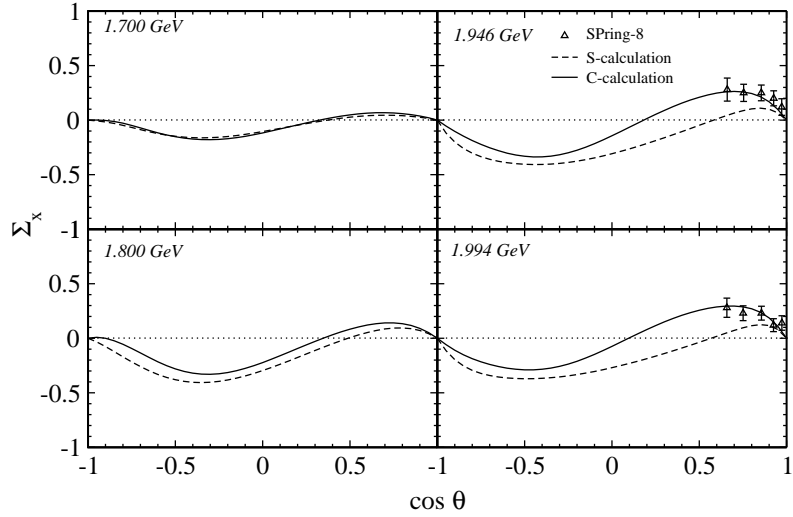}}
       \caption{
The calculated photon beam asymmetry. Data are taken from
\protect\cite{Zegers:2003}
      \label{fig7:KL}}
  \end{center}
\end{figure}

Similar to  $\pi N \to K\Lambda$ the major difference between the
$S$ and $C$ solutions is the treatment of the non-resonant
contributions. As seen in Fig.~\ref{fig5:KL}, both calculations show
two peak structures in the differential cross sections at 1.7 and
1.9~GeV. In both cases the first bump at 1.67~GeV is produced by the
$S_{11}(1650)$ resonance. The relative contributions to the second
peak at 1.9~GeV are different in the $C$ and $S$ solutions. In the
$C$-calculations this structure is described by the $S_{11}$ partial
wave. At higher energies the $S_{11}$ channel is dominated by the
non-resonant reaction mechanisms and there is no need to include a
third $S_{11}$ resonance, as done e.g. in \cite{JuliaDiaz:2005}. The
$P_{13}$ partial wave is entirely driven by the $P_{13}(1720)$ and
$P_{13}(1900)$ resonance contributions. Switching off these
resonance couplings to $K\Lambda$ leads to an almost vanishing
$P_{13}$ partial wave. In the $S$-calculations no peaking behavior
is found in the $S_{11}$ partial wave at 1.95~GeV. However, the
non-resonant effects in the $S_{11}$ channel are still important.
The role of the $P_{13}$ resonances are slightly enhanced in the
$S$-calculations. The effect from the $P_{11}(1710)$ resonance is
found to be small in both calculations due to destructive
interference with the background process. There are no significant
contributions from the spin-$\ffh$ resonances to the $\gamma N \to K
\Lambda$ reaction.

\begin{figure}
  \begin{center}
{\includegraphics*[width=4.1in]{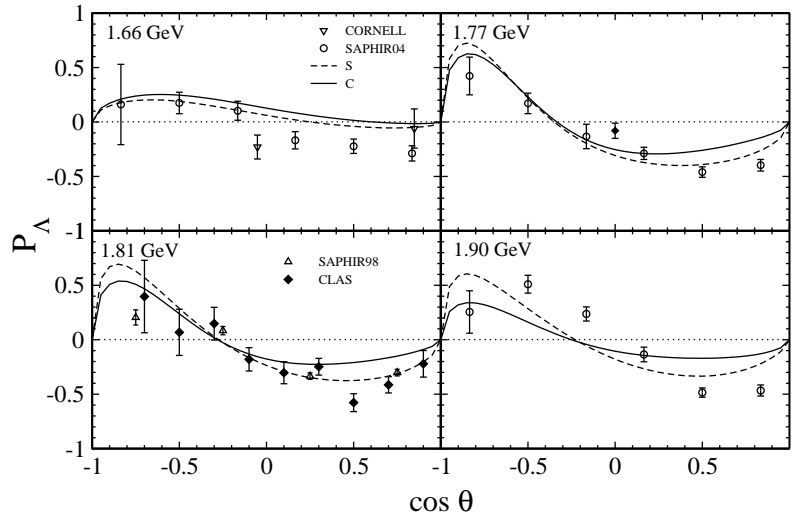}}
       \caption{
$\Lambda$-polarization in the $\gamma p \to K^+\Lambda$ reaction.
Data are from  SAPHIR98 \protect\cite{Tran:1998}, SAPHIR04
\protect\cite{Glander:2003}, CLAS  \protect\cite{McNabb:2003},
CORNELL \protect\cite{Cornell:1963}.
      \label{fig8:KL}}
  \end{center}
\end{figure}

The calculated photon beam asymmetry  $\Sigma_x$  and  recoil
polarization $P_\Lambda$ are shown in Fig.~\ref{fig7:KL} and
Fig.~\ref{fig8:KL}. Since the beam asymmetry  data from the SPring-8
collaboration \cite{Zegers:2003} are available only for energies
above 1.94 GeV, these measurements give  an insignificant constraint
on the model parameters. Therefore, the results for  the asymmetry
might be regarded as a prediction rather than an outcome of the fit.
More information comes from the $\Lambda$-polarization data. A good
description of the $\Sigma_x$ and $P_\Lambda$ data is possible in
both the $C$ and $S$ calculations.

\section{Summary and Outlook}\label{sec:sumry}

The importance of a controlled treatment of channel coupling for a
quantitative understanding of meson production on the nucleon was
pointed out. An approach, fulfilling the -- sometimes delicate --
balance between flexibility and generality, is given by using a
Lagrangian model in conjunction with a decent reaction theory. Such
a programm is underlying the Giessen model, describing meson
production by a coupled channels K-matrix approach, based on a
Lagrangian with phenomenological coupling constants and from
factors.

The results for $\omega$ meson production and associated strangeness
production by  $K\Lambda$ processes are convincing in their ability
to describe various experimental data, from total and differential
cross sections to spin observables. The close connection between
hadronic and photonic production channels was discussed for the
$\omega N$ reaction. In both $\omega N$ and $K\Lambda$ reactions the
importance of a dynamical treatment of the reaction mechanism as in
the Giessen model was evident by the fact that come of the spectral
structures were due to quantum mechanical interference phenomena. In
order to resolve those effects also in the experimental data,
measurements of spin observables play a crucial role.

\end{document}